\documentclass[aps,prl,preprint,superscriptaddress,floatfix,12pt]{revtex4-1}

\usepackage{mathrsfs}
\usepackage{graphicx}
\usepackage{amsmath}
\usepackage{amsfonts}
\usepackage{amssymb}
\usepackage{amsthm}
\usepackage{textcomp}
\usepackage{color}
\usepackage{geometry}
\usepackage[normalem]{ulem}
\usepackage{placeins}
\usepackage{bm}
\usepackage{wasysym}
\usepackage{titlesec}
\usepackage{txfonts}
\usepackage{graphicx}
\usepackage{subfigure}
\usepackage{braket}
\usepackage{verbatim}

\newcommand\R{\mathbb{R}}

\newcommand\Z{\mathbb{Z}}

\newcommand\I{\mathcal{I}}

\newcommand\Dn{\widetilde{D}}

\begin{document}

\title{A plane wave study on the localized-extended transitions in the one-dimensional incommensurate systems}

\author{Huajie Chen}
\affiliation
{School of Mathematical Sciences, Beijing Normal University, Beijing 100875, China}

\author{Aihui Zhou}
\affiliation
{LSEC, Institute of Computational Mathematics and Scientific/Engineering Computing, Academy of Mathematics and Systems Science, Chinese Academy of Sciences, Beijing 100190, China}
\affiliation
{School of Mathematical Sciences, University of Chinese Academy of Sciences, Beijing 100049, China}

\author{Yuzhi Zhou}
\affiliation
{Software Center for High Performance Numerical Simulation, China Academy of Engineering Physics, Beijing 100088, China}
\affiliation
{Institute of Applied Physics and Computational Mathematics, Beijing 100088, China}

\email[Send correspondence to:]{zhou\_yuzhi@iapcm.ac.cn}

\date{\today}

\begin{abstract}
Based on our recently proposed plane wave framework, we theoretically study the localized-extended transition in the one dimensional incommensurate systems with cosine type of potentials, which are in close connection to many recent experiments in the ultracold atom and photonic crystal. We formulate a propagator based scattering picture for the transition at the ground state and single particle mobility edge, in which the deeper connection between the incommensurate potentials, eigenstate compositions and transition mechanism is revealed. 
We further show that there exists a upper limit of localization length for all localized eigenstates, leading to an fundamental difference to the Anderson localization. Numerical calculations are presented alongside the analysis to justify our statements. The theoretical analysis and numerical methods can also be generalized to systems in higher dimensions, with different potentials or beyond the single particle regime, which would benefit the future studies in the related fields.
\end{abstract}

\maketitle


\section{I. Introduction}
\label{sec:introduction}

The localization of quantum waves in the non-periodic potentials has aroused much research interests since Anderson's seminal paper decades ago \cite{Anderson58}. Unlike the fully disordered system, the incommensurate system, which consists of two or more periodic components but lacks overall periodicity, can exhibit localized-extended transition in 1D or 2D from the experiments of ultracold atoms \cite{Roati08,Deissler10,Bordia16} and photonic crystals \cite{Segev13,Wang2020,Lahini09}, as well as from theoretical studies \cite{Modugno08,Lahini09,zhou19}. In the ultracold-atom systems, such transitions can be further robustly controlled through adjusting the incommensurate potential and interatomic interaction strength \cite{Deissler10}, which makes them an ideal platform to simulate quantum many body effects \cite{Schreiber15,Lukin20}. 
Moreover, many salient spectrum and transport properties have been observed in the incommensurate systems of 2D materials, for instance the quantum Hall effect \cite{Dean2013}, the greatly enhanced carrier mobility \cite{Kang2017}, and the unconventional superconductivity \cite{Cao18sc}. Their occurrence might deeply relate to the localization of electrons near the Fermi level \cite{Naik2018}. Therefore, a full knowledge of the incommensurate localization mechanism in the single particle regime is a prerequisite to gain a better control of the quantum states in experiments, as well as to understand related quantum many body effects and novel electronic properties.

Given the feasibility of describing the localized states, a majority of the theoretical studies on the incommensurate localization are based on the tight-binding model \cite{AAmodel,Carr20,Cazeaux19,Roscilde08,Modugno08,Madsen13,Sun2015,massatt17,Li2017i,li2020}, which greatly improve our understanding and helps to interpret related experimental results. 
However, one has to be careful in constructing the model Hamiltonian, as the oversimplification might lead to incorrect localization properties. An example is the Aubry-Andr\'{e} (AA) 1D tight binding model \cite{AAmodel}, which showed that the eigenstates are either all localized or all delocalized, depending on the relative strength between the incommensurate cosine modulation and the primary lattice. 
Yet, it has been verified in recent experiments that there exists a single-particle mobility edge (SPME) in such incommensurate systems \cite{lschenPRL}. 
Meanwhile, the existence of mobility edge can be recovered in theoretical calculations using the model Hamiltonian with more continuum nature \cite{Sun2015,Li2017i,Settino2017}, suggesting an overlook of high-order hopping effect in the AA model \cite{Sun2015,Li2017i}. 
In addition, some tight binding calculations are performed with finite size or periodic boundary condition. This would cause some troubles in distinguishing a truly localized state and an extended state but exhibiting a localized wave packet in the range of system size, which might undermine our understanding on the transition mechanism.

On the other hand, plane wave basis has several features that would benefit the study of transition in the incommensurate system. First, it is very convenient in representing the eigenstates of kinetic energy operator and the incommensurate potential, which does not require further approximations to describe the Hamiltonian in the single particle regime. Meanwhile, it is naturally compatible with extended systems and one can further circumvent the periodic boundary condition utilizing the ergodicity as discussed in \cite{zhou19}. Therefore some systematic errors from the inappropriate boundary conditions can be avoided. 
Furthermore, since the plane waves are generally viewed as the conjugate of the localized orbitals, one could expect gaining complementary perspectives on the localized-extended transition under this representation, which helps to complete our understanding on the subject. 
However, previously plane wave studies are limited due to a lack of rigorous mathematical treatment of the corresponding quantum eigenvalue problem.

In this paper, we will study the localized-extended transition of the time-independent Schr\"{o}dinger equation for the one dimensional incommensurate systems, utilizing our recently developed plane wave framework \cite{zhou19}. Specifically, we formulate a scattering picture to describe the localized-to-extended transition based on the propagation of plane waves in the higher dimension reciprocal space, without explicitly solving the eigenvalue problem. 
Here we mainly study two cases: (a) the ground state transition with increasing potential strength, and (b) the transition at the SPME, in which the deeper connection between the incommensurate potentials, the plane wave components in the eigenstates and the transition mechanism is revealed. We further discuss the existence of a maximum localization strength, which implies an intrinsic difference from the Anderson localization. (Other fundamental differences between the incommensurate localization and Anderson localization have also been discussed in recent theoretical studies \cite{Li2017i,Albert10}.) Numerical calculations under the same framework, which directly solve the eigenvalue problem, are performed to justify the conclusions from the scattering picture.
We stress that even though part of the conclusions in this paper can be drawn from some revised tight binding models, the plane wave studies provide us unique insights on the mechanism of transition. Also note that although we restrict our discussions to the incommensurate systems in one dimension with cosine-type potentials, our plane wave representation and scattering picture can in principle be used to study general incommensurate systems.

The rest of this paper is organized as follow. In Section II, we briefly introduce the plane wave framework for incommensurate systems. In Section III, we formulate the scattering picture and apply it to study the emergence of localization transition in the ground state and the localized-to-extended transition at SPME. In Section IV, we discuss the existence of a maximum localization length and compare the incommensurate localization with Anderson localization in this context. In Section V, we present some concluding remarks. Moreover, we discuss the role of incommensurate ratio on the transition based on the scattering picture in the Appendix.

\section{II. Plane wave framework}
\label{sec:planewave}

In this section we briefly introduce the plane wave framework for the simulations of the incommensurate systems.
We consider the following time-independent Schr\"{o}dinger equation for an one-dimension incommensurate system with two periodic components:
\begin{equation}
\label{eq:eigen}
\left( -\frac{1}{2}\frac{\partial^2}{\partial x^2} + V_1(x) + V_2(x) \right)~\Psi(x) = E~\Psi(x) \quad \forall~x\in\R,
\end{equation}
where $V_1$ and $V_2$ are periodic potentials
$V_j(x + n\tau_j)=V_j(x)$ with $n\in\Z$ and $\tau_j$ the lattice constants for $j=1,2$.
The incommensurateness puts further constraints on the ratio between $\tau_j$ and the corresponding reciprocal lattice $G_j=2\pi/\tau_j$, that $\frac{\tau_2}{\tau_1} = \frac{G_1}{G_2} = \beta$ is an irrational number.
This leads to the so-called ergodicity (see Section II.2) and is crucial to the discussions in this work.

\subsection{II.1 The plane wave discretization}
\label{subsec:pw}

Following the discussions in \cite{zhou19}, we use the basis functions $\big\{e^{i(mG_1+nG_2)x}\big\}_{(m,n)\in\I_{E_{cut}}}$ with index set
\begin{eqnarray}
\I_{E_{cut}}:=\left\{(m,n)\in\Z^2:~|mG_1|^2+|nG_2|^2\leq 2 E_{cut}\right\}
\end{eqnarray}
to discretize the Schr\"{o}dinger equation \eqref{eq:eigen}, where $ E_{cut}$ is the energy cutoff that features the accuracy and computational cost of this discretization.
The ground state wave function $\Psi(x)$ is approximated by
$\Psi_{E_{cut}}(x) = \sum_{(m,n)\in\I_{E_{cut}}}u_{mn}e^{i(mG_1+nG_2)x}$
with $\hat{u}=\{u_{mn}\}_{(m,n)\in\I_{E_{cut}}}$ the unknown coefficients.
Eq. \eqref{eq:eigen} is then discretized into a matrix eigenvalue problem
\begin{eqnarray}
\label{eq:eigen-matrix}
H\hat{u}=E\hat{u},
\end{eqnarray}
where the hamiltonian matrix elements are given by
\begin{align}
\label{H_elements}
H_{mn,m'n'}(k) = \frac{1}{2} \big|G_{1m}+G_{2n}\big|^2 \delta_{mm'}\delta_{nn'} + V_{1(m-m')}\delta_{nn'} + V_{2(n-n')}\delta_{mm'}
\qquad (m,n),(m',n')\in\I_{E_{cut}}
\end{align}
with $V_{jm}$ the Fourier component of the periodic potential
$V_{jm} = \frac{1}{\tau_j}\int_0^{\tau_j}V_j(x)e^{-imG_j x} {\rm d}x$.

To quantitatively describe the extent of localization for a wavefunction $\Psi_{E_{cut}}$, it is convenient to use the inverse participation ratio (IPR) \cite{Kramer1993} to measure the number plane waves contributing a given eigenstate, which is defined by
\begin{equation}
\label{eq:IPR}
{\rm IPR}\big(\Psi_{E_{cut}}\big) := \sum_{(m,n)\in\I_{E_{cut}}} \big|u_{mn}\big|^4~.
\end{equation}
For an extended state, its IPR will scale like $O(1)$ as $E_{cut}\rightarrow\infty$.
While for a localized state, the IPR will be approaching $0$ with the scaling $O\big(E_{cut}^{-1/2}\big)$
as $E_{cut}\rightarrow\infty$. The scaling factor $1/2$ will be discussed later.

\subsection{II.2 Ergodicity and the higher dimensional interpretation}
\label{subsec:ergodicity}

We will discuss the concept of ergodicity of incommensurate systems particularly in the higher dimensional representation. We refer to \cite{zhou19} for more details.

The ergodicity was originally used to describes the equiprobable access to all states in the phase space in thermodynamics.
The ergodicity in our context is a direct consequence from the incommensurateness, and is the root of many unique properties of the incommensurate systems.
It can be stated in the mathematical language as those in Ref. \cite{cances17,dingzhou09,massatt17}.
Here we prefer a more direct description:
with infinitely large cutoffs $E_{cut}$, the coupled wave vectors $\{mG_1+nG_2\}_{m,n\in\Z}$ will fill the whole reciprocal space $\R$ \emph{densely, uniformly} and \emph{unrepeatedly}.

The one dimensional Schr\"{o}dinger equation \eqref{eq:eigen} can be reformulated in $\R\times\R$ by
\begin{eqnarray}
\label{eq:eigen-high}
\left( -\frac{1}{2}\Dn + V_1(x) + V_2(x') \right)~\tilde{\Psi}(x,x') = E~\tilde{\Psi}(x,x')
\qquad\forall~(x,x')\in \R\times\R
\end{eqnarray}
with the directional derivative
$\Dn~\tilde{\Psi}(x,x') :=  \big(\frac{\partial}{\partial x} + \frac{\partial}{\partial x'}\big)^2~\tilde{\Psi}(x,x')$.
Since the potential $V_1(x) + V_2(x')$ is periodic in $\R\times\R$, the periodicity is restored by this higher dimensional representation. We note that similar idea has been explored for describing the lattices and diffraction patterns of quasi-crystals (see e.g. Ref.~\cite{baake17,blionv15,jiang14,Walter09}).
It has been shown in \cite{zhou19} that, with an energy cutoff $E_{cut}$ and the basis set $\big\{e^{i(mG_1x+nG_2x')}\big\}_{(m,n)\in\I_{E_{cut}}}$, the discretization of Eq. \eqref{eq:eigen-high} leads to the same matrix eigenvalue problem Eq. \eqref{eq:eigen-matrix} at $\Gamma$ point.
The solution in Eq.~\eqref{eq:eigen-high} can further be transformed back to that of Eq.~\eqref{eq:eigen} by taking the diagonal $\Psi(x) = \tilde{\Psi}(x,x)$.

\begin{figure}[htb!]
	\includegraphics[height=7.5cm]{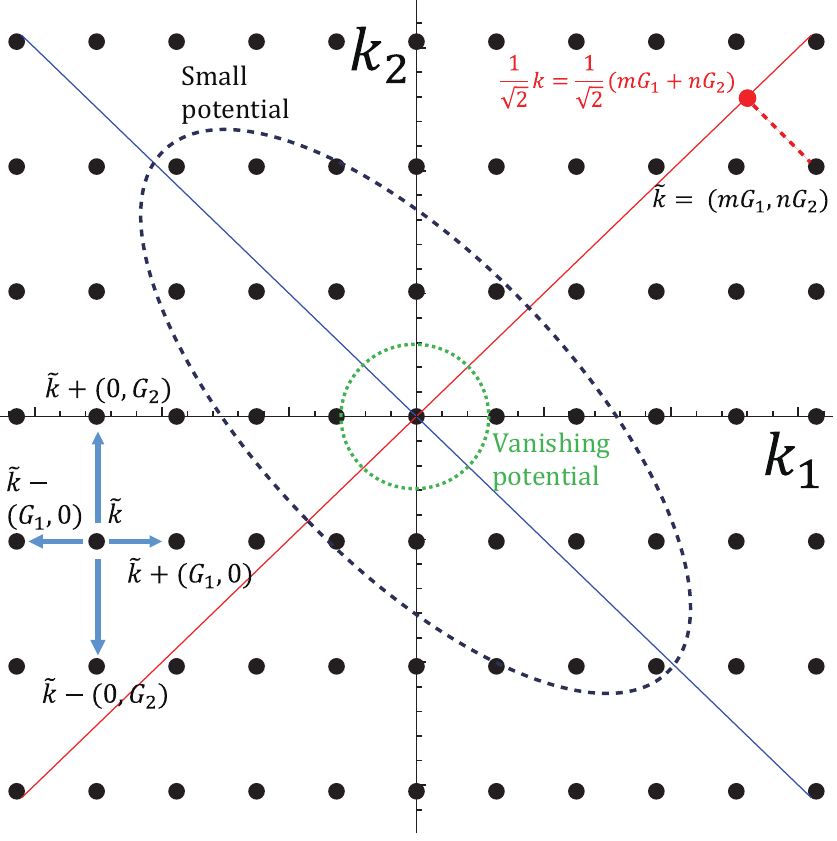}
	\caption{
		The higher dimension representation of the coupled plane waves. This 2D lattice has periodicity $\beta$ (with the ratio $\beta=\frac{\sqrt{5}-1}{2}$) in $k_1$ direction and $1$ in $k_2$ direction.
        Upper right: A 2D lattice site $\tilde{\pmb{k}}=(mG_1,nG_2)$ is projected to an 1D wave vector $\pmb{k}=mG_1+nG_2$.
		Lower left: the cosine-type potentials in the Hamiltonian scatter $\ket{\pmb{k}}$ to its nearest neighbor states.
		Central: At vanishing $V$, the ground eigenstate is mainly composed of the plane waves near the origin (green dotted circle).
		With increasing potential strength, more plane waves along $\langle -1,1 \rangle$ direction are involved to the ground state (black dashed ellipsoid). Note there is a factor of $\frac{1}{\sqrt{2}}$ from the projection on the line $k_1 = k_2$ in 2D reciprocal space. Be aware that this factor is not present in the transformation from 2D to 1D reciprocal space.
	}
	\label{fig:scatter}
\end{figure}

We can also observe the ergodicity by the projection in higher dimensional reciprocal space, as illustrated in the upper right of Fig.~\ref{fig:scatter}.
When a wave vector $\tilde{\pmb{k}}=(mG_1,nG_2)~(m,n\in\Z)$ on the two dimensional reciprocal lattice is projected onto line $k_1 - k_2 = 0$, it gives the one dimensional wave vector $\pmb{k}=mG_1+nG_2$.
The ergodicity is reflected by the fact that all the projected points will densely, uniformly and unrepeatedly spread on the line $k_1 - k_2 = 0$.
This observation is crucial to the discussions in the following.

\section{III. The scattering picture of localized-extended transition}
\label{sec:localization-extension}

In this section, we will formulate the transition picture within the plane wave framework.
For simplicity, we will restrict ourselves to the following incommensurate Hamiltonian in the discussions of localized-extended transition,
\begin{equation}
\label{eq:HamilCos}
\hat{H} = -\frac{1}{2}\frac{\partial^2}{\partial x^2} + V_1\cos\big(\beta x\big) + V_2\cos\big(x\big)
\end{equation}
with $\beta$ an irrational number and $V_1,V_2>0$ the strengths of potentials.
This type of incommensurate potential are commonly used in many ultracold atom experiments \cite{Fallani07,Roati08,Deissler10} and the photonic crystals \cite{Lahini09,Wang2020}.
Note that Eq.~\eqref{eq:HamilCos} is a special case of the Hamiltonian in Eq.~\eqref{eq:eigen} with $G_1=\beta$ and $G_2=1$ and $V_1(x)$ and $V_2(x)$ being cosine potential.
The discussions in this section can be extended to more general potentials.

It is convenient to rewrite the Hamiltonian in the second quantization form:
\begin{equation}
\label{eq:second}
\hat{H} = \sum_{\pmb{k}=(mG_1,nG_2),~m,n\in\Z} \frac{|\pmb{k}|_{\#}^2}{2} c_{\pmb{k}}^\dagger c_{\pmb{k}} + \sum_{\pmb{k}} \left(V_1 c_{\pmb{k}}^\dagger c_{\pmb{k}+(G_1,0)} + V_2 c_{\pmb{k}}^\dagger c_{\pmb{k}+(0,G_2)} + h.c. \right) ,
\end{equation}
where $c_{\pmb{k}}^+$ and $c_{\pmb{k}}$ are the annihilation and creation operator associated with the plane wave state $|\pmb{k}\rangle$.
It is important to note that the norm $|\cdot|_{\#}$ in \eqref{eq:second} for a plane wave $\pmb{k}=(mG_1,nG_2)$ is given by $|\pmb{k}|_{\#}:=|\pmb{k}_1+\pmb{k}_2|=|mG_1+nG_2|$, rather than the standard Euclidean norm $\big(|nG_1|^2+|mG_2|^2\big)^{1/2}$.
This is essentially a tight-binding Hamiltonian with nearest neighbor hopping in the two dimensional reciprocal lattice, as shown in the lower left of Fig.~\ref{fig:scatter}.

In the following, we first qualitatively investigate the change of plane wave components in the ground state as the potential strength grows.
Then we formulate a scattering picture using the language of propagator, which is further adopted to study the transitions at the ground state and SPME.
For simplicity we only consider the $V_1 = V_2 = V$ case in this section. The discussion of $V_1 \neq V_2$ case is presented in Appendix.

\subsection{III.1 Transition at the ground states: a qualitative study}
\label{sec:qualitative}

At the vanishing potential strength $V$, one expects the ground state of $\hat{H}$ (defined in \eqref{eq:second} with $V_1=V_2=V$) is mainly composed of the plane waves near the origin. As $V$ increases, more plane waves are mixed into the ground state as the coupling to nearby sites becomes more significant. This is illustrated in Fig.~\ref{fig:scatter}, in which the green dotted circle grows to the black dashed ellipsoid as $V$ increases. The ellipsoid shape in the figure can be understood by the following arguments.
The kinetic (on-site) energy of a site $\pmb{k}=(mG_1,mnG_2)$ is $\frac{1}{2}|\pmb{k}|^2_{\#} = \frac{1}{2}|mG_1+nG_2|^2$, which grows much faster along $\langle 1, 1 \rangle$ direction but remains fluctuated around certain value along $\langle 1, \bar{1} \rangle$ direction.
Thus we expect less plane waves along $\langle 1, 1 \rangle$ direction, whose on-site energies are significantly higher than the ground state energy, to be mixed into the ground state compared with those along $\langle 1,\bar{1} \rangle$ direction.
For now, the ground state solution mainly consists of a finite number of plane waves and is an extended wave function in the real space. Consequently, the corresponding IPR value (defined in Eq.~\eqref{eq:IPR}) mainly depends on the distribution in the bounded circle or ellipsoid and will not decay as the energy cutoff $E_{cut}$ increases.

\begin{figure}[htb!]
	\includegraphics[height=6.6cm]{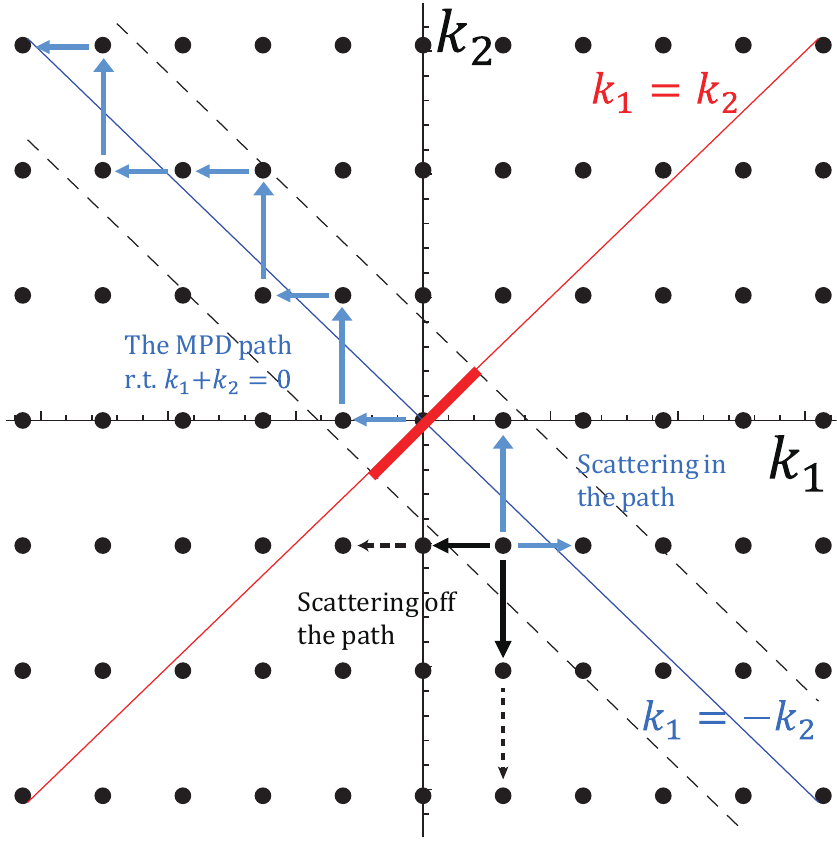}
	\caption{An MPD path (blue arrows) in the 2D reciprocal space that connects the origin and the lattice sites along $\langle 1, \bar{1} \rangle$ direction. The lower half is the same by symmetry.
	The anti-diagonal dashed lines indicates the boundary of the path.
	The lattice sites on this path, when transformed back to the 1D reciprocal space (by projecting onto line $k_1=k_2$), form a continuous band around the origin (red thick bar).
	The dashed black arrows represent the scattering events to states off the path, whose probability amplitude decays fast away from the boundary.}
	\label{fig:path}
\end{figure}

If $V$ further increases and crosses the critical point, then the ellipsoid of the plane wave components will become a "stripe" that extends to infinity along $\langle 1,\bar{1}\rangle$ direction.
Given the form of Eq.~\eqref{eq:second}, there must exist some "scattering" paths that connect the all plane waves close to $k_1+k_2=0$, through nearest neighbor hopping.
As will be quantitatively formulated later, the most relevant path to the transition is the one with all sites staying at closest distance to the line $k_1 + k_2 = 0$ to maximize the hopping probability, as depicted in Fig.~\ref{fig:path}.
We will call them "the most probable direct path" (MPD path) respect to $k_1 + k_2 = 0$ in the following (the MPD paths respect to other anti-diagonal lines $k_1 + k_2 = k$ are also relevant to the discussion of SPME).
In this case, there are infinitely many plane waves, which are connected by the paths extending to infinity, markedly contributing to the ground state. If we further project all the involved plane waves onto the line $k_1 - k_2 = 0$, they form a continuous (more precisely, densely distributed) band around the origin in one dimensional reciprocal space due to the ergodicity (see also Fig.~\ref{fig:path}). 
Now the interval between the $k$ points in the 1D reciprocal space now becomes zero, and the ground state undergoes a localization transition.  This observation is similar to the discussion of quasi particle lifetime and the localization of Green's function in the time domain in \cite[Chapter 3 and Appendix H]{mattuck1976}.  In addition, since the distribution of plane waves is semi 1D along $k_1 + k_2 = 0$, the IPR value vanishes as $O\big(E_{cut}^{-1/2}\big)$, as we have mentioned in Sec.~II.

\begin{figure}[htb!]
 	\includegraphics[width=4.9cm]{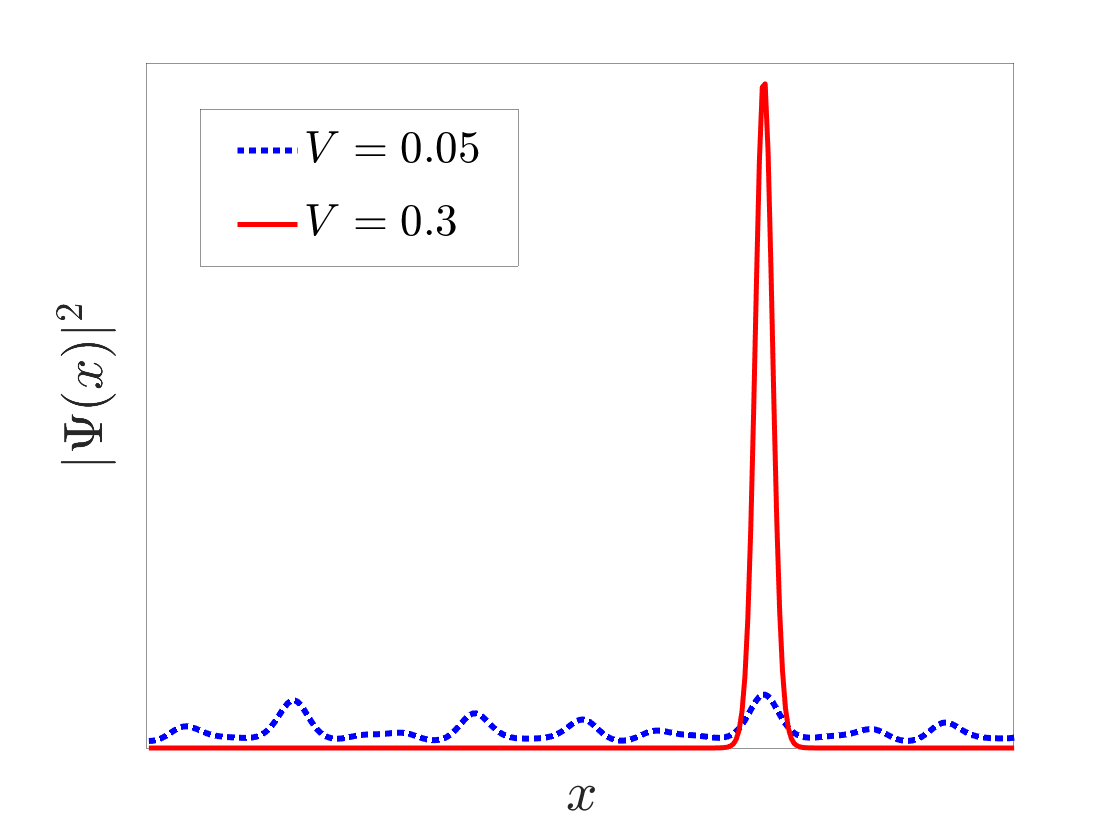}
 	\includegraphics[width=4.9cm]{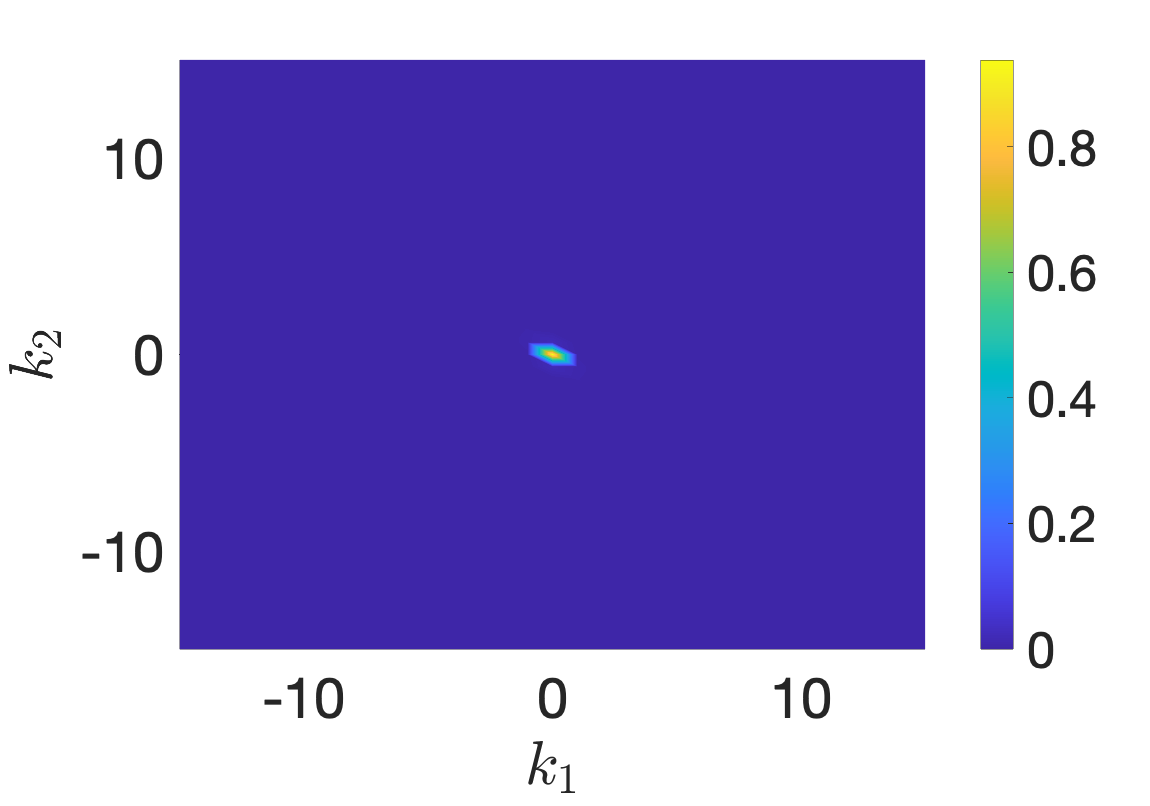}
 	\includegraphics[width=4.9cm]{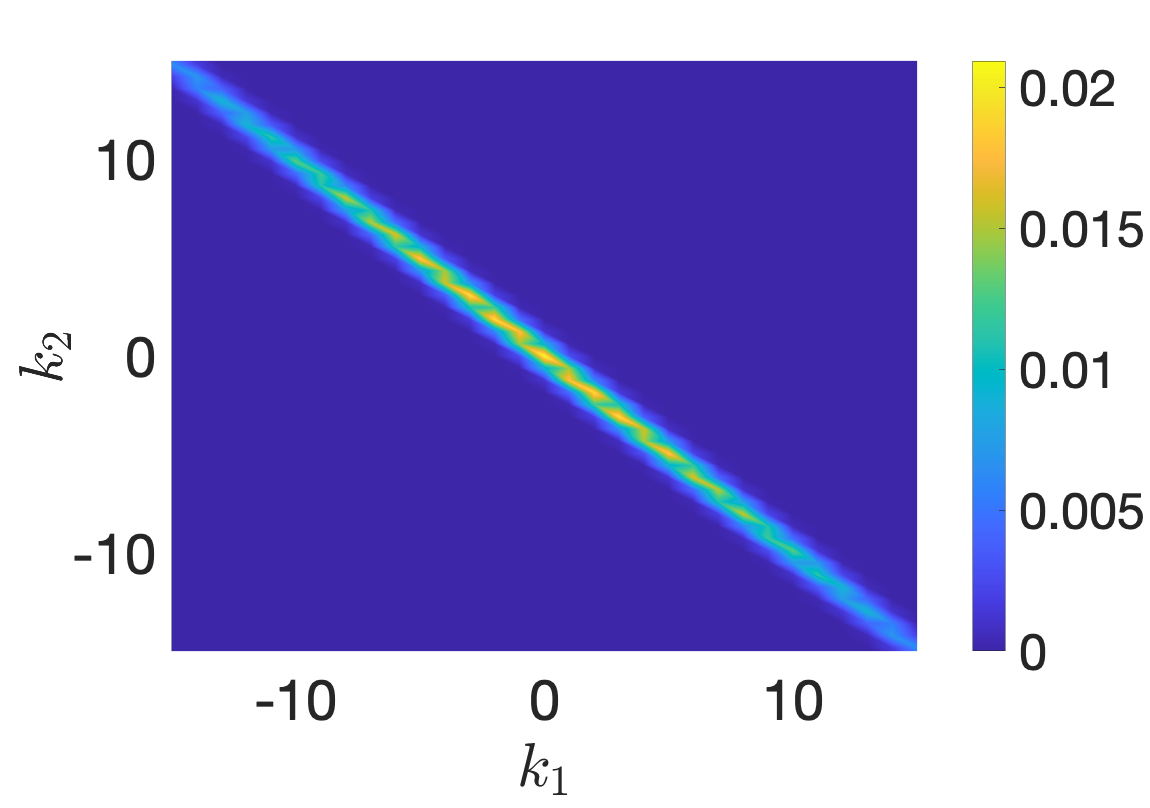}
 	\caption{\label{fig:calsca} The square of ground state solutions in 1D real space and 2D reciprocal space for two systems with $V = 0.05$ and $V = 0.3$ respectively.
 	As the potential strength increases, the ground state becomes more localized in real space and more extensive along $\langle \bar{1},1 \rangle$ direction in reciprocal space.
 	}
 \end{figure}

We then verify the above statements by numerical calculations (see \cite{zhou19} for details of the algorithm).
We take the ratio $\beta=\frac{1}{2}\big(\sqrt{5}-1\big)$, and simulate two incommensurate systems with $V=0.05$ and $V=0.3$, corresponding to extended state and localized state respectively (these potential strengths are consistent with the critical strength derived at Sec.~III.3 ). Their ground state solutions are compared in Fig.~\ref{fig:calsca}. 
With stronger potential, we observe a much more extensive distribution of the occupied plane waves along the anti-diagonal direction, which is consistent with the above analysis.

\subsection{III.2 A propagator-based formulation}
\label{sec:scatter}

The above picture can be translated into a propagator-based formulation that enables the quantitative study of the transition, without explicitly solving the eigenvalue problem.
For simplicity of notations, we will denote by $\pmb{k}=(mG_1,nG_2)$ the state of a plane wave $e^{i(mG_1+nG_2)x}$, and $E_{\pmb{k}}=\frac{1}{2}|\pmb{k}|^2_{\#}$ the corresponding kinetic energy.

The propagation of a plane wave $\pmb{k}_i$ being scattered once by the potential to its neighbor $\pmb{k}_j$ has the probability amplitude:
\begin{equation}
T(\pmb{k}_i \rightarrow \pmb{k}_j,E) = \frac{1}{E-E_i}~V~\frac{1}{E-E_j}~,
\end{equation}
with $E_i = E_{\pmb{k}_i}$, $E_j = E_{\pmb{k}_j}$, and $E$ the frequency of the free propagator.
Then the probability amplitude for an $N$ successive scattering events along a path $\mathcal{P}=\big\{\pmb{k}_0,\cdots,\pmb{k}_N\big\}$ is given by:
\begin{equation}
\label{T:path}
T(\mathcal{P},E) = \frac{1}{E-E_0}~V~\frac{1}{E-E_1}~V~\cdots~V~\frac{1}{E-E_N} ~.
\end{equation}
For a MPD path $\mathcal{P}_{k}$ respect to $k_1+k_2=k$, we take $N\to\infty$ in Eq.~\eqref{T:path}, which reads
\begin{equation}
\mathcal{T}(\mathcal{P}_{k},E) =  \lim_{N \to \infty} \prod_{i=1}^{N} \frac{V}{E-E_i}~.
\end{equation}

Intuitively, $T(\mathcal{P}_{k},E)$ can be viewed as a term in the diagram expansion for the Green's function of the Hamiltonian Eq.~\eqref{eq:second}, and one could solve for the Green's function to retrieve the properties of the system in principle.
This is exactly what Anderson did in his paper: he adopted the Renormalized Perturbation Expansion (RPE) of the Green's function \cite{Feenberg1948,Watson1957} to study the localization in the disordered systems \cite{Anderson58}.
The analysis was later simplified by Ziman \cite{Ziman1969}, Thouless \cite{Thouless1970}, and further reorganized by Economou and Cohen \cite{Economou1970,Economou1972}.
The Anderson's original formulation is very complicated, and one could imagine that the it gets even more complicated in the incommensurate systems since they are essentially higher dimensional problems.

In this paper, we will not go into full details of the Green's function expansions when studying the extended-to-localized transition. Instead, at the edge of the transition, one finds some specific MPD paths only just connect to the plane waves at infinity, and the corresponding $T(\mathcal{P},E)$ undergoes an abrupt change from 0 to nearly divergence or the other way around. This marks the situation we have mentioned previously: the infinite number of plane waves start or cease to markedly mix into the eigenstate, which leads to the localized-extended transition at the ground state or mobility edge. Therefore, we investigate the divergence criterion of $T(\mathcal{P},E)$ which represents the abrupt change of the plane wave distribution of the eigenstates during the transition. The reason for choosing the MPD paths is that they can more effectively reach the plane waves at infinity than other paths, thus are most relevant to the transition.

Before we quantitatively describe this situation, one more subtlety needs to be considered. Given the 2D nature of the problem, the plane wave states within the MPD path are inevitably
scattered off the path into nearby states, as depicted in Fig.~\ref{fig:path}. This creates other non-MPD paths to infinity by allowing for short digressions from the MPD path. This results in a reduced probability to reach the infinity for the original MPD path.
Since it could happen at any site within the MPD path, we multiply a factor $\alpha\in(0,1)$ for each scattering event. ($\alpha$ can also be seen as the average effect from the non-MPD paths in the diagram expansion.)
Then the probability amplitude reads:
\begin{equation}
\label{eq:revpath}
\mathcal{T}(\mathcal{P}_{k},E) =  \lim_{N \to \infty} \prod_{i=1}^{N} \frac{\alpha V}{E-E_i}~.
\end{equation}
Thus, at the transition point it should fulfill
\begin{equation}
\label{eq:crit}
\frac{\alpha V}{\overline{|E-E_i|}} \approx 1,
\end{equation}
at given $E$ or $V$, where $\overline{|E-E_i|}$ indicates the geometric mean over all sites in the path $\mathcal{P}$. 

The above analysis sacrifices a little bit of rigorousness but facilitates our understanding with a more direct physical picture. This idea is similar to the analysis proposed by Ziman on the Anderson localization \cite{Ziman1969}. The conclusions drawn from this scattering picture will be further checked by the numerical calculations in the same plane wave framework, where the full information of the eigenpairs is obtained.

\subsection{III.3 Transition at the ground states: a quantitative study}
\label{sec:quantitative}

Now we adopt the scattering picture to quantitatively describe the localization transition of the ground
state. Before the potential strength $V$ reaching the critical point $V_{\rm c}$, the $T(\mathcal{P},E)$ for any MPD path makes negligible contribution to the diagram expansion for any frequency $E$.
This is because $V<V_{\rm c}$ makes $\frac{\alpha V}{\overline{|E-E_i|}} < 1$, hence the infinite multiplication in $\mathcal{T}(\mathcal{P},E)$ makes it an infinitesimal. This is consistent with previous analysis and calculations that the plane waves infinitely away make inappreciable contribution to the ground state, which then corresponds to an extended wavefunction.

To calculate the critical point $V_{\rm c}$ from Eq.~\eqref{eq:revpath}, we consider the MPD path $\mathcal{P}_{0}$ respect to $k_1+k_2=0$, since this path has the smallest site energies on average thus can minimize the denominator in Eq.~\eqref{eq:crit}. Its site-averaged natural logarithm of $\mathcal{T}(\mathcal{P}_{0},E)$ is given by
\begin{equation}
\label{eq:Delta1}
\Delta(\mathcal{P}_{0},E) = \lim_{N\to\infty} \frac{1}{N} \sum_{k_n \in \mathcal{P}_{0}} \ln \frac{\alpha V}{\Big|E-\frac{1}{2}|\pmb{k}_n|_{\#}^2\Big|} ~.
\end{equation}
From the ergodicity, the projected plane waves form a uniform and continuous band centering the origin (see the red thick bar in Fig. \ref{fig:path}). Thus we can transform the summation in Eq.~\eqref{eq:Delta1} into an integration:
\begin{equation}
\label{eq:Delta2}
\Delta(\mathcal{P}_{0},E) = \frac{1}{G_1+G_2} \int_{-\frac{G_1+G_2}{2}}^{\frac{G_1+G_2}{2}}\ln\frac{\alpha V}{\big|E-\frac{1}{2}s^2\big|} ~{\rm d} s ~ ,
\end{equation}
where $s$ corresponds to the norm $|\pmb{k}|_{\#}$.
The integral region is determined as follow. The hopping along $\mathcal{P}_0$ in Fig.~\ref{fig:path} proceeds with alternative and balanced upward and leftward jumps to retain the closest distance to $k_1 + k_2=0$, in which process the maximum distance from $k_1 + k_2 = 0$ is $\frac{G_1+G_2}{2}$. This gives a boundary of $[-\frac{G+Q}{2},\frac{G+Q}{2}]$.

Now we make a rough estimate of the factor $\alpha$.
First, $\alpha$ is not close to $0$ since the plane wave energies change quadratically away from the path, thus the major contribution to the ground state still comes from the plane waves within the path, which have significantly lower energies.
Second, the scattering to the plane waves right near the path is not negligible. They have comparable energies thus can be effectively included into the ground state by other slightly detouring paths.
As a consequence, one would expect a considerable transfer of the amplitude to the nearest neighbor plane waves of the MPD path. Therefore $\alpha$ can not be close to 1 either.
Then $\alpha = 1/2$ could be a reasonable choice and we will use it in the following calculations.

To solve for the critical potential strength $V_c$, we further parameterize $E = 0$ to obtain a minimal value of $V_{\rm c}$. From the criterion Eq.~\eqref{eq:crit}, we have
\begin{equation}
\Delta(\mathcal{P}_{0},0)= 2 + \ln \frac{8 \alpha V_{\rm c}}{\big(G_1+G_2\big)^2} \approx 0~.
\end{equation}
Plugging in $\alpha = 1/2$, $G_1 = \beta=\frac{1}{2}\big(\sqrt{5}-1\big)$ and $G_2 = 1$, we can obtain the critical value $V_{\rm c} = 0.089$ for our exemplified system.

\begin{figure}[htb!]
	\includegraphics[width=7.0cm]{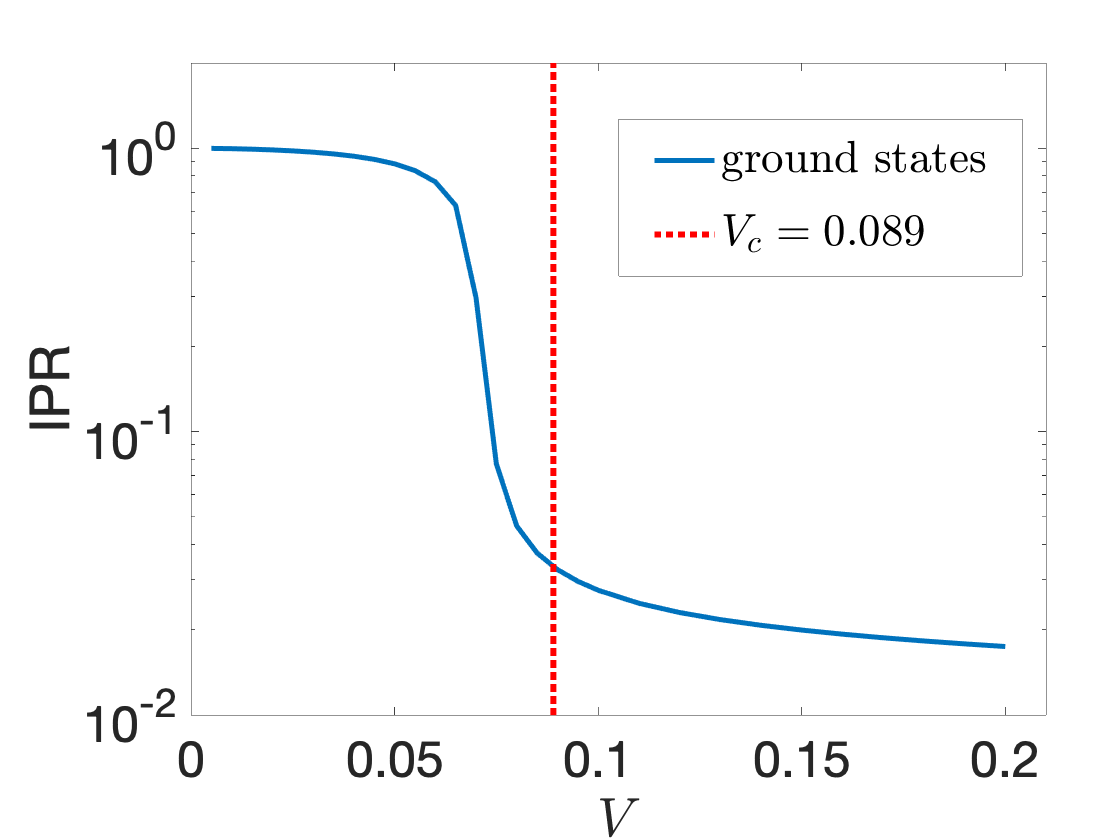}
	\caption{The IPR of the ground state with varying potential strengths.
	The predicted critical strength $V_{\rm c} = 0.089$ (indicated by the red dashed line) is in good agreement with the numerical results.}
	\label{fig:iprgs}
\end{figure}

To numerically verify the results from the scattering picture, we compute the ground states with the ratio $\beta=\frac{1}{2}\big(\sqrt{5}-1\big)$ and varying potential strengths $V$ and plot the corresponding IPR values in Fig.~\ref{fig:iprgs}. We observe that the slope of IPR changes significantly around $V = 0.08 \sim 0.10$, indicating the occurrence of the transition in this region. This is in quantitative agreement with our estimated value $V_{\rm c}=0.089$.

\subsection{III.4 Transition at the mobility edge}
\label{subsec:mobility}

\begin{figure}[htb!]
	\includegraphics[width=6.6cm]{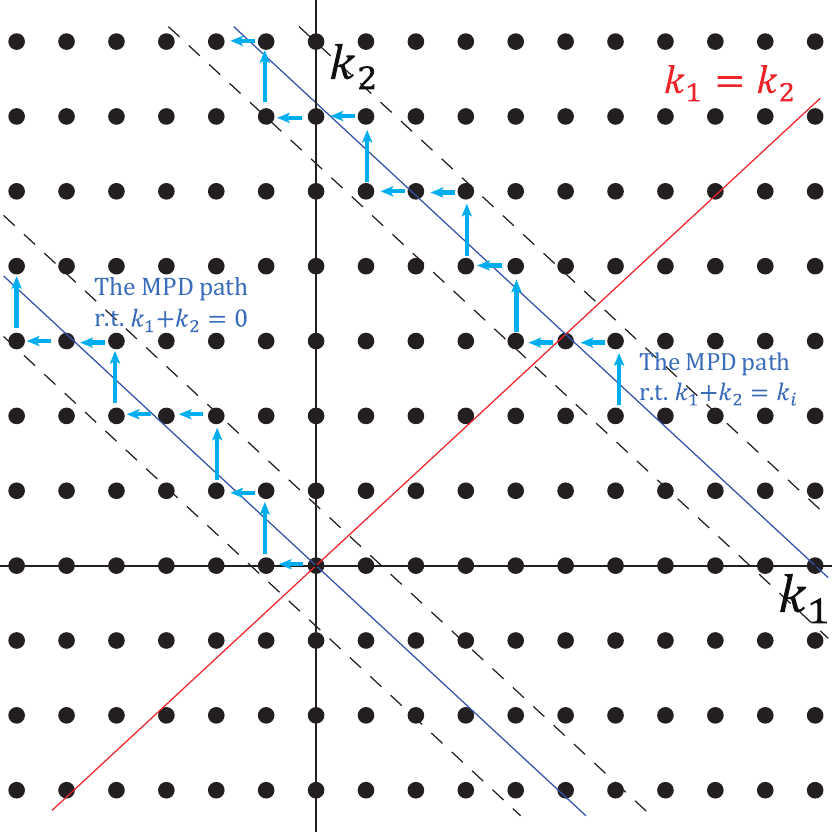}
	\caption{The MPD paths (blue arrows) respect to different $k$.
	These two paths can be translated from one to the other by a direct shift due to the ergodicity.
	}
	\label{fig:twopath}
\end{figure}

For the transition at SPME, we need to consider the situation with $V>V_{\rm c}$.
Since $V$ has crossed the critical point, not only $\mathcal{P}_{0}$ for the ground state satisfies $\Delta(\mathcal{P}_{0},E) \geq 0$,
there also exist MPD paths $\mathcal{P}_{k}$ and higher frequency $\tilde{E}$ such that the condition $\Delta(\mathcal{P}_{k},\tilde{E}) \geq 0$ can also hold.
Note that the MPD paths respect to different $k$ can exactly overlap with each other by a translational shift (see Fig.~\ref{fig:twopath}).
This is due to the ergodicity of incommensurate systems:
if there is a site $\pmb{k}$ in the MPD path respect to $k_x + k_y = k_1$, then due to the ergodicity, there exists another site $\pmb{\tilde{k}}$ whose relative position to $k_x + k_y = \tilde{k}_1$ is arbitrarily close to that between $\pmb{k}$ and $k_x + k_y = k_1$. Hence shifting $\pmb{k}$ to $\pmb{\tilde{k}}$ will overlap the two paths.

Despite their same geometries, the MPD paths with respect to larger $|k|$ have higher energy differences between neighboring sites, which in general reduces the overall probability amplitude. This further results in fewer paths that connect to $k$ points at infinity. 
Then there exists a critical path $\mathcal{P}_{\pm kc}$ respect to $k_1 + k_2 = \pm k_c$ (together with a corresponding frequency $E_c$) such that the MPD paths with higher $|k|$ can not connects to infinity, which indicates the onset of localized-to-extended transition and sets the SPME.

We shall estimate the critical energy $k_c$ using the scattering picture. For simplicity, we assume the sites of $\mathcal{P}_{k_c}$ stay on the one side of $k_1 + k_2 = 0$. Then $E_c$ can be parameterized as $\frac{1}{2}\big(k_c - \frac{G_1+G_2}{2}\big)^2$ to minimize the denominator, which gives largest $k_c$. We can obtain $k_c$ by solving
\begin{equation}
\label{eq:me}
\Delta(\mathcal{P}_{k_c},E_c) = \frac{1}{G_1 + G_2} \int_{k_c-\frac{G_1+G_2}{2}}^{k_c+\frac{G_1+G_2}{2}}\ln\frac{\alpha V}{\big|\frac{1}{2}\big(k_c - \frac{G_1+G_2}{2}\big)^2-\frac{1}{2}s^2\big|}~{\rm d} s = 0,
\end{equation}
where the integral region $[k_c-\frac{G+Q}{2}, k_c+\frac{G+Q}{2}]$ is derived by similar analysis as that for \eqref{eq:Delta2}. The estimation of $\alpha$ is also similar to previous section, except now the plane wave energies increase quadratically in one direction away from the path while decrease quadratically in the other.
We stress that the $E_{\rm c}$ is {\it not} the SPME, but defines the upper and lower bounds for the plane wave components of the eigenstate near the SPME.
This is because $E_{\rm c}$ is the frequency of the free propagator or loosely regarded as the unperturbed energy, while the SPME is defined respect to the solved eigenspectrum.

\begin{figure}[htb!]
	\includegraphics[width=7.5cm]{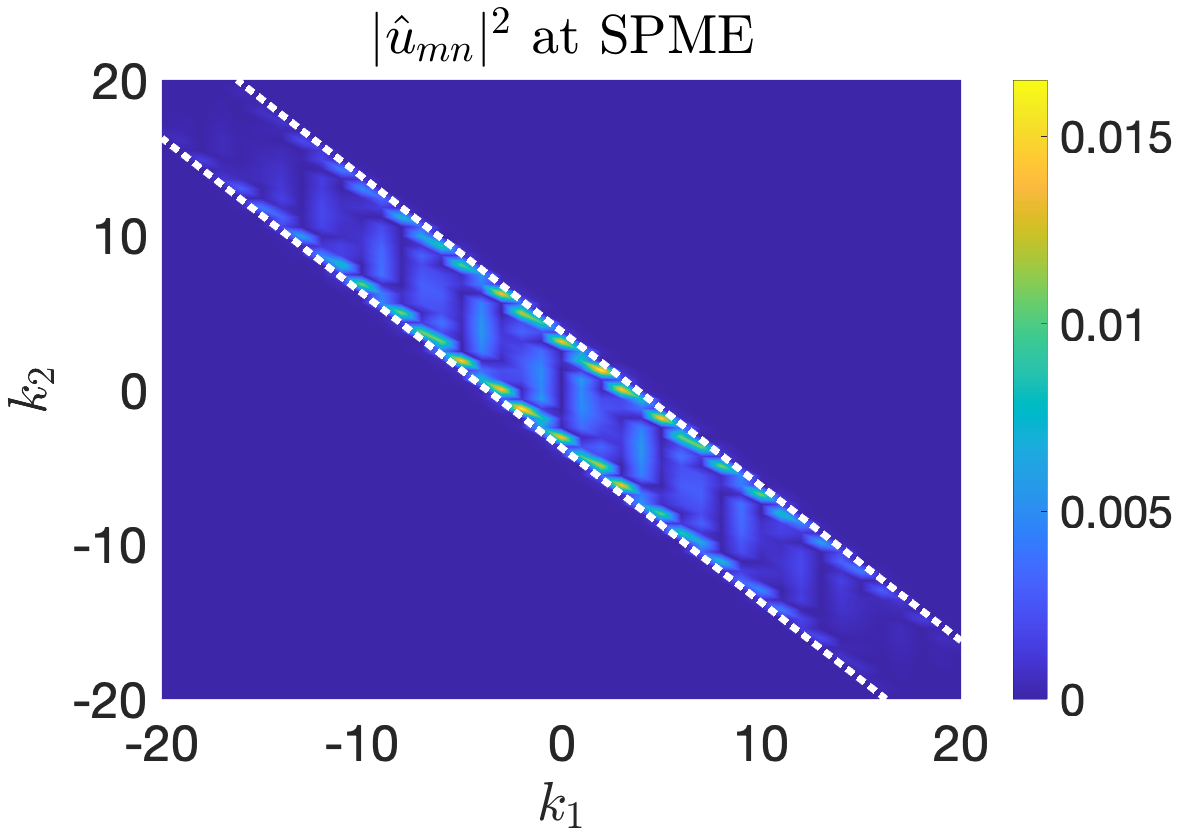}
	\caption{The localized state with highest energy in 2D reciprocal space.
	The dashed lines indicate the predicted boundary of the reciprocal space distribution.}
	\label{fig:Ec}
\end{figure}

As an example, we set the potential strength $V = 3.0$ and estimate corresponding $k_c$ using the scattering picture. Plugging in $\alpha = 1/2$, $G_1 = \beta$ and $G_2 = 1$ and integrating Eq.~\eqref{eq:me} analytically, we obtain $k_c = 2.950$. Then for the eigenstate at SPME, the boundary for the plane wave components is $[-k_c-\frac{G+Q}{2}$, $k_c+\frac{G+Q}{2}]$ (note the lower bound is obtained by the symmetry).
In Fig.~\ref{fig:Ec}, we present the region $|k_1+k_2|\leq k_c+\frac{G+Q}{2}$, together with the eigenstate at SPME in the reciprocal space.
The eigenstate at SPME is obtained by solving the eigenvalue problem Eq.~\eqref{eq:eigen} by plane wave methods and searching through the whole eigenspectrum.
We observe in the figure that our theoretical prediction of the boundary matches well with the numerics.

Moreover, the existence of the SPME goes against the prediction from the AA tight binding model, which states that the eigenstates are either all localized or all delocalized, determined by the ratio between the strengths of the primary and secondary lattices \cite{AAmodel}.
The origin of the discrepancy can be understood by the fact that the AA Hamiltonian is in principle a single band model under extreme tight binding limit and cannot represent the properties of the full spectrum in more general cases.
We further emphasis that the localization properties are independent of the basis set used to discretize the Hamiltonian.
For this reason, the SPME has also been observed in some revised tight binding models \cite{Sun2015,li2020}, real space calculations \cite{Settino2017,Li2017i}, and plane wave calculations in this paper.

\section{IV. Comparison with the Anderson localization}
\label{sec:compare-anderson}

One obvious difference between the incommensurate localization and Anderson localization is the existence of SPME, which has already been discussed in the existing theoretical and experimental literature \cite{Li2017i,lschenPRL}.
In this paper, we focus on a new implication from the scattering picture in the higher dimensional reciprocal lattice, which lead to another fundamental difference on the localization length.

Assuming negligible inelastic scattering and infinite sample size for simplicity, the localization length in the Anderson localization of 1D disordered system in principle could be arbitrarily large. This can be illustrated using Anderson's analysis \cite{Anderson1980} based on Landauer's conductance formula \cite{Landauer1970}. 
The key feature in the Anderson localization is that the intermediate state between two successive scattering of the conducting particle (generally taken as the plane wave states for simplicity) is random. The average over possible intermediate states leads to a linear dependance of $\ln (1+S(L))~(\sim \gamma L)$ on $L$, where $S$ is the dimensionless resistance for a length $L$ of the sample and $\gamma$ the linear factor. 
This, in return, gives rise to the exponential decay of the wavefunction in the real space and $1/\gamma$ determines the localization length. It can be further envisioned that by tuning the mean free path through the defect concentration, one could in principle have arbitrarily small $\gamma$, hence arbitrarily large localization length.

While for the incommensurate localization, it is a different picture: the intermediate state between two scattering events is fixed.
This means the mechanism leading to Anderson localization will not apply for the incommensurate system:
unlike the previous case where a continuum of intermediate states are visited by defect-average in a single event of propagation, now it is achieved by \emph{the infinite number} of scattering events along the MPD paths in the reciprocal space.
Consequently, the minimum width of the paths constrains the upper limit of localization length in the real space.
In other words, a localized wavefunction with very large localization length in the real space would require the distribution of the plane waves to behave more or less like ``delta function", where the continuum of the $k$ only exists in a very narrow region.
However, this is against the scattering picture, where the continuum of projected wave vectors is achieved by some MPD paths that have a minimum width of $G_1+G_2$.

\begin{figure}[htb!]
	\includegraphics[width=10.0cm]{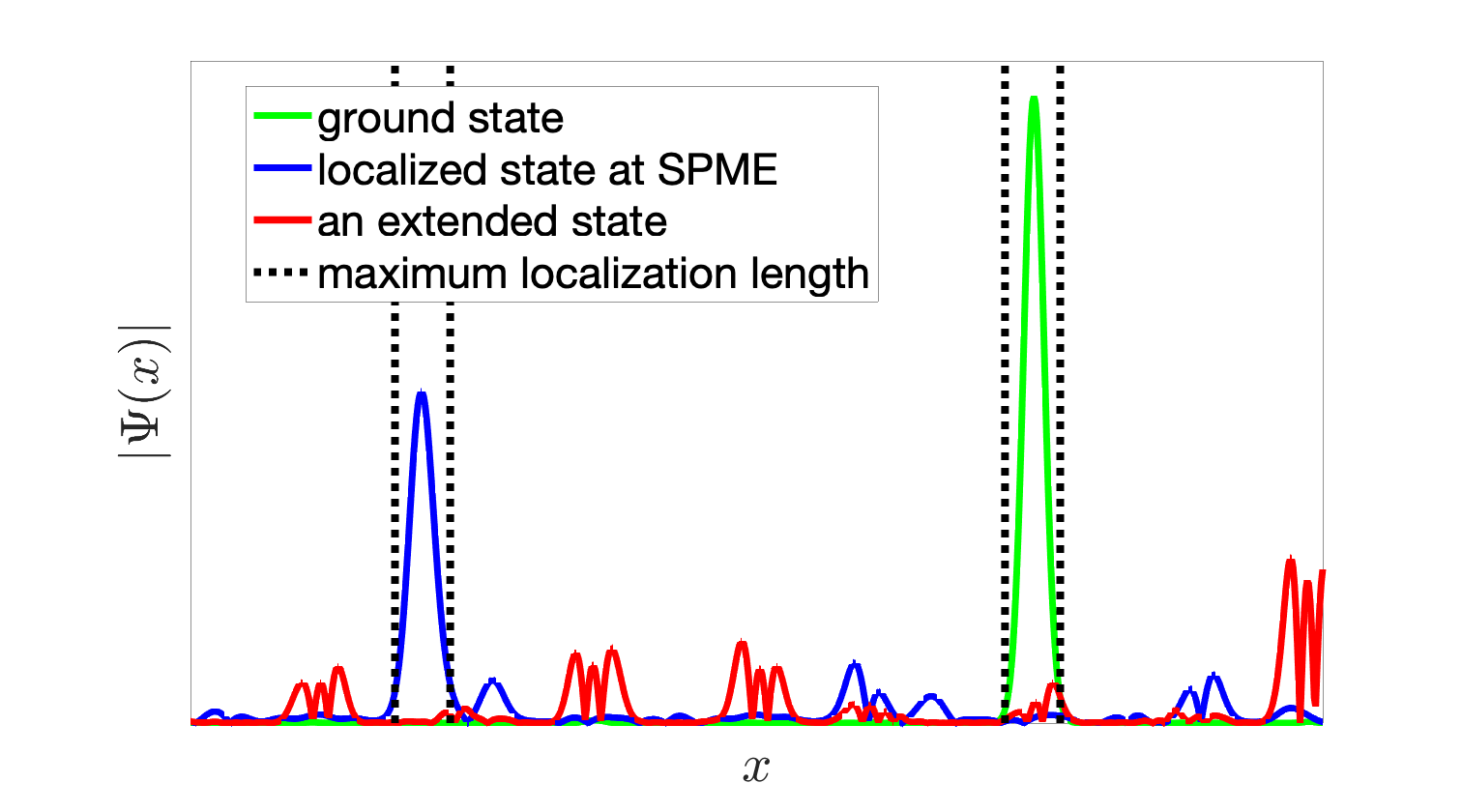}
	\caption{The ground state, the highest localized state and an extended state in the real space with $V = 3.0$. The vertical lines indicate the maximum localization length from our analysis.}
	\label{fig:length}
\end{figure}

We can further illustrate this fact by numerical simulations for the system with $V = 3.0$.
We plot the norm of wavefunction for the ground state, the highest localized state and an extended state in the real space in Fig.~\ref{fig:length}.
From the figure, the localized states exhibit a localization length on the scale of $2\pi/(1+\beta)$, though there exists minor contribution outside the major localization region.

Given this difference, the attempts of using incommensurate systems to simulate the Anderson localization might require further justification. However, we also note that the finite size effect and noises in the potential might blur the boundary between these two types of localization. Distinguishing incommensurate localization and Anderson localization in experiments thus seems to be an interesting and challenging question for future studies.

\section{V. Conclusions}
\label{sec:conclusions}

In this paper, we utilize the plane wave framework to study the extended-to-localized transitions in the 1D incommensurate systems.
A scattering picture has been formulated to quantitatively study the transitions of at the ground states and SPME.
Under this picture, we further discuss the fundamental difference between the incommensurate localization and Anderson localization.
The numerical calculations have been conducted alongside to justify the conclusions from the scattering picture.
In principle, the theoretical analysis and numerical methods can be carried over to more general incommensurate systems in higher dimensions, with more complicated form of potentials and beyond the single particle regime, thus provide theoretical tools to investigate spectrum and transport properties of the incommensurate systems in various fields.

\section{Acknowledgements}
This work was partially supported by National Key Research and Development of China under grant 2019YFA0709601.
H. Chen's work was also partially supported by the Natural Science Foundation of China under grant 11971066.
A. Zhou's work was also partially supported by the Key Research Program of Frontier Sciences of the Chinese Academy of Sciences under grant QYZDJ-SSW-SYS010, and the National Science Foundation of China under grant 11671389.

\section{Appendix. Role of the incommensurate ratio}
\label{sec:appendixV12}

The ratio $\beta$ between the periodicity of the periodic components is the key feature of the incommensurate system.
But in many existing works on the localization, the role of this value has not been fully explored.
In this appendix, we will investigate the extended-to-localized transition with respect to the ratio $\beta$.
We will not restrict ourselves to the case $V_1=V_2=V$ as in Section III, but consider general systems that allow $V_1 \neq V_2$.

For the general cases, the probability amplitude of a path $\mathcal{P}$ at frequency $E$ in Eq. \eqref{T:path} is
\begin{equation}
T(\mathcal{P},E) = \frac{1}{E-E_0}\cdot V_{q(1)}\cdot\frac{1}{E-E_1}\cdot V_{q(2)}\cdot\frac{1}{E-E_2}~\cdots~V_{q(N)}\cdot\frac{1}{E-E_N}~,
\end{equation}
where $q(i)=1$ if the hopping $\pmb{k}_{i-1}\rightarrow\pmb{k}_i$ is parallel to the $k_1$ direction and $q(i)=2$ if the hopping is parallel to the $k_2$ direction.
For the MPD path, the number of the horizontal ($k_1$ direction) and vertical ($k_2$ direction) jumps has a ratio of $1/\beta$ since the path is extending to infinity along $\langle \bar{1},1 \rangle$ direction.
With this observation, we can rewrite Eq. \eqref{eq:Delta2} as
\begin{equation}
\Delta(E)
= \frac{1}{G_1+G_2} \int_{-\frac{G_1+G_2}{2}}^{\frac{G_1+G_2}{2}}\ln\frac{V_1^{\frac{1}{1+\beta}}\cdot V_2^{\frac{\beta}{1+\beta}}\cdot \alpha}{\big|E-\frac{1}{2}s^2\big|} ~{\rm d} s  .
\end{equation}
For the same critical condition that $\Delta(E) \approx 0$, we have
\begin{equation}
\label{eq:ratio}
\Delta(E=0) = 2 + \ln\frac{8 V_1^{\frac{1}{1+\beta}}\cdot V_2^{\frac{\beta}{1+\beta}}\cdot\alpha}{G_2^2~(1+\beta)^2} \approx 0,
\end{equation}
where we have replaced $G_1$ by $\beta G_2$ in this equation to better illustrate the role of $\beta$.
We see from Eq. \eqref{eq:ratio} that the incommensurate ratio $\beta$ influences the transition in two places.
First, in the nominator of the propagator, it controls the weight of each periodic components in the geometric mean of the potential strength.
Second, in the denominator of the propagator, it reflects the energy differences between plane wave states connected by the incommensurate potential.
We note Eq. \eqref{eq:ratio} can provide guidelines to the manipulation of the localization transition in experiments of ultracold atoms and photonic crystals.

\begin{figure}[htb!]
	\includegraphics[width=7.0cm]{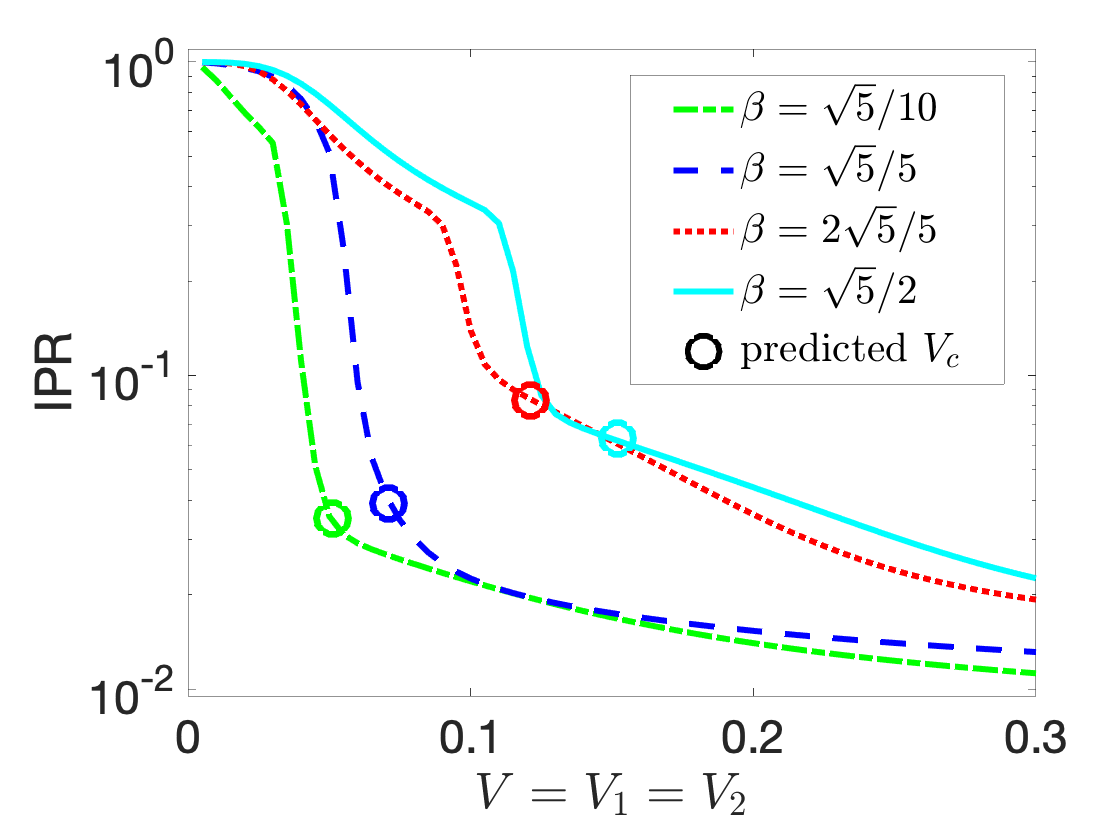}
	\includegraphics[width=7.0cm]{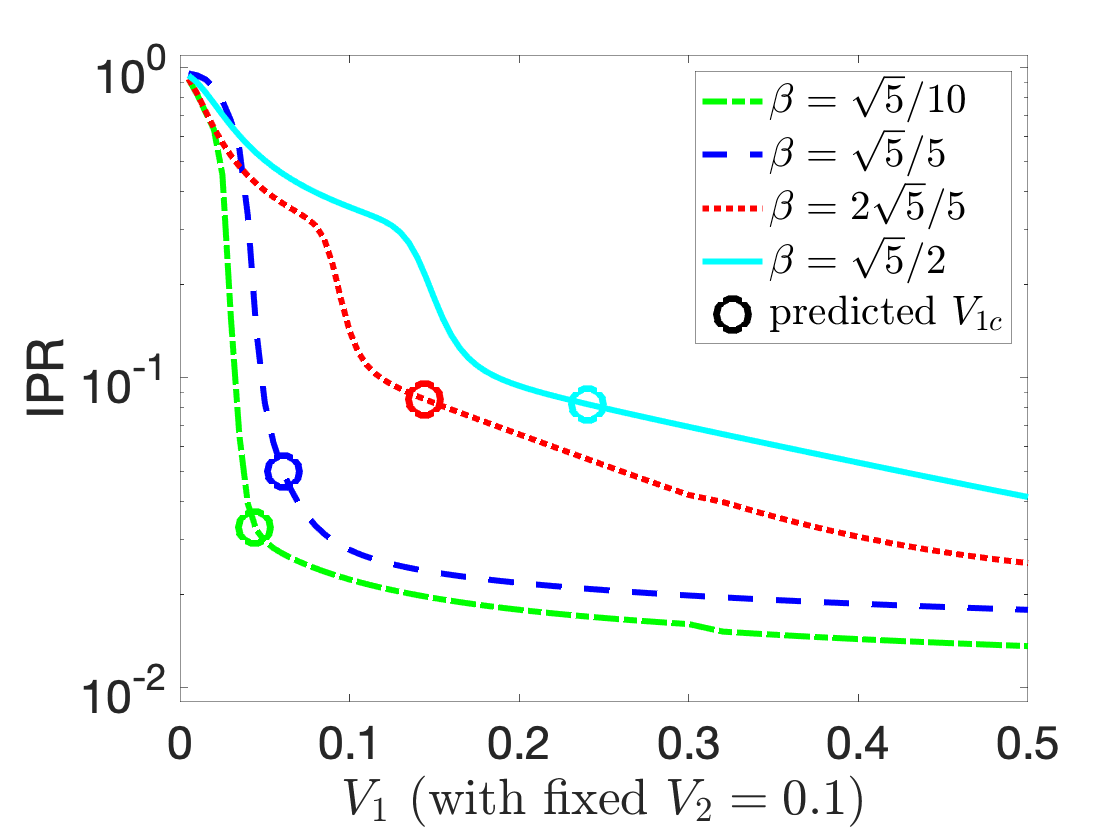}
	\caption{(a) The IPR varied $V=V_1= V_2$ for different ratios $\beta$. (b) The IPR with fixed $V_2 = 0.1$ and varied $V_1$ for different ratios $\beta$.
	The predicted critical value $V_{\rm c}$'s in both scenarios are shown by circles.}
	\label{fig:beta}
\end{figure}

In the following, we demonstrate two simple scenarios of such manipulations through numerical calculations.
In the first one, we have $V_1 = V_2 = V$ and the incommensurate ratio $\beta$ is varied ranging from $0.1 \sim 1.1$, which is a direct extension of the case in Sec.~III.
This can be achieved by varying the wavelength of the laser while keeping the intensity fixed in the ultracold atom experiments.
The IPR plots for systems with different $\beta$ are shown in Fig.~\ref{fig:beta} (a), together with the predicted critical potential $V_c$'s using Eq. \eqref{eq:ratio} (shown by circles in same color).
We can see in all cases the predicted $V_c$'s are in good agreement with the trends of IPR from numerical calculations.
In the second scenario, we fix $V_2 = 0.1$ and check how the critical value of $V_1$, denoted by $V_{1c}$, changes with $\beta$.
This is a modeling of the experimental setup when the primary periodic lattice is fixed, while the incommensurate modulation from the secondary lattice are varied.
The corresponding IPR plots and the predicted $V_{1c}$ from Eq. \eqref{eq:ratio} are shown in Fig.~\ref{fig:beta} (b). Again we see good agreement between $V_{1c}$ and the trend of IPR.
These numerical simulations further justify the scattering picture for the transitions.

\bibliography{zhou}

\end{document}